\documentclass[sigconf]{acmart}

\renewcommand\footnotetextcopyrightpermission[1]{} 

\AtBeginDocument{%
  }

\usepackage{amsmath}
\usepackage{graphicx}
\usepackage{booktabs}
\usepackage{xcolor}
\usepackage{comment}

\usepackage{algorithm}
\usepackage{algpseudocode}

\begin{document}


\title{Making Strong Error-Correcting Codes Work Effectively for HBM in AI Inference}


\author{Rui Xie}
\orcid{0000-0003-3177-5071}
\affiliation{%
  \institution{Rensselaer Polytechnic Institute}
  \city{Troy}
  \state{New York}
  \country{USA}
}
\email{xier2@rpi.edu}

\author{Yunhua Fang}
\orcid{0009-0009-4718-8825}
\affiliation{%
  \institution{Rensselaer Polytechnic Institute}
  \city{Troy}
  \state{New York}
  \country{USA}
}
\email{fangy6@rpi.edu}

\author{Asad Ul Haq}
\orcid{0009-0003-7975-0102}
\affiliation{%
  \institution{Rensselaer Polytechnic Institute}
  \city{Troy}
  \state{New York}
  \country{USA}
}
\email{asadul@rpi.edu}

\author{Linsen Ma}
\orcid{0009-0000-8535-7911}
\affiliation{%
  \institution{ScaleFlux}
  \city{Milpitas}
  \state{California}
  \country{USA}
}
\email{linsen.ma@scaleflux.com}

\author{Sanchari Sen}
\orcid{0000-0003-0080-2882}
\affiliation{%
  \institution{IBM T.J. Watson Research Center}
  \city{Yorktown Heights}
  \state{New York}
  \country{USA}
}
\email{Sanchari.Sen@ibm.com}

\author{Swagath Venkataramani}
\orcid{0000-0002-0470-6364}
\affiliation{%
  \institution{IBM T.J. Watson Research Center}
  \city{Yorktown Heights}
  \state{New York}
  \country{USA}
}
\email{Swagath.Venkataramani@ibm.com}

\author{Liu Liu}
\orcid{0000-0003-0792-8146}
\affiliation{%
  \institution{Rensselaer Polytechnic Institute}
  \city{Troy}
  \state{New York}
  \country{USA}
}
\email{liu.liu@rpi.edu}

\author{Tong Zhang}
\orcid{0009-0009-8005-0043}
\affiliation{%
  \institution{Rensselaer Polytechnic Institute}
  \city{Troy}
  \state{New York}
  \country{USA}
}
\email{zhangt4@rpi.edu}

\begin{abstract}
LLM inference is increasingly memory-bound, and HBM cost per GB now dominates system cost. Today’s HBM stacks include short on-die ECC, which tightens binning, raises price, and locks reliability policy inside the device. This paper asks a simple question: can we tolerate a much higher raw HBM bit error rate (BER) and still keep end-to-end correctness and throughput, without changing the HBM PHY or the fixed 32\,B transaction size? We propose \textbf{REACH} (Reliability Extension Architecture for Cost-effective HBM), a controller-managed ECC design that keeps the HBM link and 32\,B transfers unchanged. REACH uses a two-level Reed--Solomon (RS) scheme: each 32\,B chunk uses an inner RS code to check and correct most faults locally, while chunks that cannot be fixed are marked as erasures. An outer RS code spans kilobytes and runs in erasure-only mode, repairing only the flagged chunks and avoiding the expensive locator step. For small random writes, REACH updates outer parity with differential parity so it does not recompute parity over the whole span, and an optional importance-adaptive bit-plane policy can protect only critical fields (for example, BF16 exponents) to reduce ECC work and traffic. On three LLMs at 8K context, REACH keeps \(\sim\)79\% of the on-die ECC throughput at BER=0 and stays qualified up to raw BER \(=10^{-3}\), extending tolerable device error rates by about three orders of magnitude while keeping tokens/s nearly flat. In ASAP7, a full REACH controller occupies 15.2\,mm$^2$ and consumes 17.5\,W at 3.56\,TB/s (\(\sim\)4.9\,pJ/byte), and it reduces ECC area and power by 11.6$\times$ and \(\sim\)60\% compared to a naive long-RS baseline. By moving strong ECC into the controller, REACH turns long-code reliability into a system choice, enabling vendors to trade higher device BER for lower HBM \$/GB under the same standard interface.
\end{abstract}





\maketitle

\section{Introduction}

High-Bandwidth Memory (HBM) has become a foundational element of modern AI systems because it is the only commodity memory that can feed today’s accelerators at multi-TB/s within a practical energy budget. Large AI models stream tens to hundreds of gigabytes of weights with high reuse, so end-to-end inference throughput is often gated by memory bandwidth. The downside is cost: HBM remains 5–10$\times$ more expensive than conventional DRAM in \$/GB~\cite{Koch2024TheMW}, and that premium now dominates total system cost for inference clusters.

There are only two levers to push against this cost wall. One can reduce the number of bits stored in HBM via compression, quantization, or sparsification~\cite{wan2023efficient,frantar2022gptq, xie2025amplifying, alizadeh2024llm}, or one can reduce the manufacturing cost of HBM itself. The first lever has seen substantial progress but is ultimately bounded by accuracy targets and software compatibility. Even with ideal compression, the system still pays for tightly binned HBM dies with fixed on-die Error-Correcting Code (ECC). This paper instead explores the second lever from a system angle: keep the HBM physical interface and bandwidth, but change where and how reliability is provided so the memory die no longer carries short, hardwired ECC and strict binning constraints.

\begin{figure}[htbp]
    \centering
    \includegraphics[width=\linewidth]{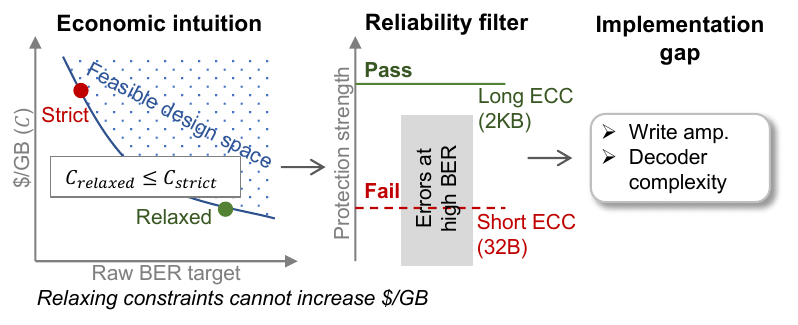}
    \caption{\textbf{Why controller-defined long ECC.}
      \textbf{Left:} Relaxing the acceptable raw BER expands the feasible design space and, under a monotone yield curve, should not increase HBM \$/GB ($C_{\text{relaxed}} \le C_{\text{strict}}$).
      \textbf{Middle:} At high raw BER (e.g., $10^{-3}$), short 32\,B ECC cannot meet reliability targets, while long-span ECC (e.g., 2\,KB) is theoretically sufficient.
      \textbf{Right:} Mapping long ECC onto a fixed 32\,B HBM interface creates an implementation gap: write amplification and decoder complexity.}
    \label{fig:problem_overview}
\end{figure}

Our economic premise is illustrated in Fig.~\ref{fig:problem_overview}: for a fixed process, \textbf{allowing a higher acceptable raw bit error rate (BER) should not increase cost per usable GB}. A relaxed BER target increases yield per wafer and reduces test/repair overhead. Today’s HBM parts use short on-die ECC to enforce a low device-level raw BER. We ask whether a system can instead sustain LLM throughput at much higher raw BER (e.g., $10^{-3}$) under the standard 32\,B HBM interface. If successful, this opens a design region where vendors can safely trade higher device noise for lower \$/GB.

To achieve this, we must shift fault tolerance from the HBM stack into the memory controller. 
Storage systems already follow this pattern: hard disk drives and SSDs tolerate relatively high raw bit error rates by using strong controller-managed ECC~\cite{zhao2013ldpc}. However, HBM introduces unique constraints. Inference workloads dominate future deployments, making the controller the natural place to implement policy-aware reliability. Yet, preserving integrity at $10^{-3}$ BER requires much stronger protection than current devices offer. Information theory dictates the use of long codewords, moving from today’s 16--32\,B granularity~\cite{gurumurthi2021hbm3} to spans of 512\,B--2\,KB. While long codes provide the necessary minimum distance (i.e., strong protection capability), applying them creates two critical system barriers:

\noindent $\bullet$ \textbf{Issue 1: Read and write amplification for small, random accesses.}
LLM inference is mostly sequential, but small random reads and writes still occur (e.g., metadata and sparse KV-cache updates). With long codewords, touching one 32\,B chunk can force a full-codeword read or read–modify–write, causing severe bandwidth amplification.

\noindent $\bullet$ \textbf{Issue 2: Decoder complexity and controller area and power.}
Long Reed–Solomon (RS) codes provide strong protection, but conventional decoders must both locate and correct errors. The locator stage scales with codeword length and field size, so a wide long-code RS decoder in a multi-TB/s HBM controller quickly becomes a major consumer of area and power.

We resolve the conflict through a controller-centric architecture called \textbf{REACH} (\textbf{R}eliability \textbf{E}xtension \textbf{A}rchitecture for \textbf{C}ost-effective \textbf{H}BM) that preserves the standard HBM PHY. REACH relies on a single decision boundary: each 32\,B chunk carries a short ``inner'' RS code that locally verifies integrity or flags the chunk as an \textit{erasure}. A long ``outer'' RS code then runs in erasure-only mode over kilobyte spans. By converting uncertain errors into known erasures at 32\,B granularity, we eliminate the expensive locator logic from the long code. Furthermore, we employ differential parity updates for random writes, which mathematically bounds amplification by touching only the target data and parity symbols. Additionally, we exploit the structure of AI data types (e.g., BF16) using a bit-plane layout~\cite{xie2024smartquant,xie2025amplifying} to selectively protect only critical fields (like exponents), further reducing overhead. This paper makes the following contributions:

\noindent $\bullet$ \textbf{Contribution 1: Controller-defined reliability.}
We frame HBM reliability as a controller-defined system resource and show that short on-die ECC behind a fixed 32\,B HBM link cannot exploit long-code distance, erasure-only decoding, or unequal bit importance.

\noindent $\bullet$ \textbf{Contribution 2: Two-level codeword with bounded amplification.}
We design the REACH architecture, which utilizes a hierarchical RS organization and differential parity to turn 32\,B chunks into known erasures for a long outer code and, by construction, bounds small-access amplification and limits each access to at most one long-span repair.

\noindent $\bullet$ \textbf{Contribution 3: Practical HBM-scale implementation.}
We demonstrate practicality by evaluating LLM inference at 8K context and a 7nm implementation, achieving about 79\% of on-die ECC throughput at BER=0, qualified operation up to raw BER=$10^{-3}$, and 11.6$\times$ area and $\sim$60\% power savings over a naive long-RS baseline at 3.5TB/s-class bandwidth.

The remainder of this paper is organized as follows. Section~\ref{sec:background} provides background on HBM reliability constraints and the cost-yield relationship. Section~\ref{sec:method} details REACH. Section~\ref{sec:verification} analytically verifies the reliability bounds and erasure probabilities of the design. Section~\ref{sec:evaluation} evaluates the experimental methodology, end-to-end performance on LLM workloads, and hardware implementation costs. Section~\ref{sec:related} discusses related work, and Section~\ref{sec:conclusion} concludes.

\section{Background and Motivation}
\label{sec:background}

\subsection{HBM Architecture and the Cost of Fixed Reliability}
\label{sec:cost-model}

HBM delivers bandwidth by stacking DRAM dies with TSVs and driving a very wide, short-reach interface. HBM3 keeps a 1024-bit aggregate datapath, organized as 16 channels (32 pseudo-channels), and transacts in fixed 32\,B packets to sustain multi-TB/s at low pJ/bit. The PHY width and the 32\,B transaction size are hard constraints set by the JEDEC HBM interface~\cite{JEDEC_JESD270_4_2025,JEDEC_JESD238B01_2025}. System designers cannot change them without breaking compatibility.

Today’s HBM parts provide protection in two \textit{built-in} layers~\cite{JEDEC_JESD270_4_2025,JEDEC_JESD238B01_2025}. First, the off-die link carries link-integrity signals per burst (parity and data-bus inversion) that are generated and checked by the controller PHY and the DRAM PHY. Second, the device implements on-die ECC with background error check and scrub, and exposes coarse severity reporting to the host. In addition, many deployed accelerators add a third, controller- or subsystem-managed ECC/RAS layer over cache lines or pages. For example, NVIDIA documents handling of HBM ECC events together with row remapping and HBM channel repair in the GPU memory subsystem~\cite{nvidia_a100_mem_err_mgmt2025,nvidia_row_remapping2023}, and the Linux EDAC stack added HBM row-retirement support for AMD Instinct MI300 platforms~\cite{larabel_linux69_mi300_edac_2024}. This third layer is platform specific and sits above the link and device mechanisms.

We distinguish two error classes. (i) Independent bit flips are a tractable lower-bound model for soft errors and background noise. They make capacity and invocation rates analytically checkable. (ii) Correlated faults are common in practice: short bursts within a few bytes, row or column defects, TSV or half-channel issues~\cite{wu2024removing}. These produce clusters inside one 32\,B unit or across adjacent units. On-die ECC masks some of these faults when the short code covers them, but guaranteeing that coverage forces vendors to ship at very low raw BER. That requires tighter process control, extra redundancy, and longer screening, all of which raise \$/GB.

We can make this cost pressure explicit with a simple yield-based view. Let $p(\theta)$ be the probability that a die meets a raw bit error rate (BER) threshold $\theta$. Let $C_{\text{wafer}}$ be the wafer cost, $N_{\text{dies}}$ the number of dies per wafer, and $\text{capacity\_per\_die}$ the nominal capacity of one die. A simple cost-per-GB model is
\begin{equation}
    C(\theta) =
    \frac{C_{\text{wafer}}}{N_{\text{dies}} \cdot p(\theta) \cdot \text{capacity\_per\_die}}.
    \label{eq: cost model}
\end{equation}
We do not assume any particular functional form for $p(\theta)$, only a weak monotonicity assumption:
\begin{equation}
    \theta_2 > \theta_1 \quad \Rightarrow \quad p(\theta_2) \ge p(\theta_1).
\end{equation}
Intuitively, if the device is allowed to have a higher acceptable raw BER, at least as many dies pass binning, usually more. Eq.~\eqref{eq: cost model} then implies $C(\theta_2) \le C(\theta_1)$: relaxing the raw-BER target cannot make cost per usable GB worse.

When reliability policy is locked inside the device, the only lever is to push the internal raw BER target $\theta_{\text{dev}}$ down through process and redundancy. That increases binning and test cost and permanently ties protection to a short code and to the 32\,B packet, regardless of how the system actually accesses memory. In contrast, if the system can tolerate a higher device-level raw BER and close end-to-end integrity in the controller, the effective threshold becomes a system choice. The controller-defined ECC scheme in this paper keeps LLM tokens/s nearly flat up to raw BER $\theta_{\text{sys}} \approx 10^{-3}$ under the same JEDEC PHY and 32\,B transactions. For any monotone $p(\cdot)$, Eq.~\eqref{eq: cost model} then guarantees
\begin{equation}
    C(\theta_{\text{sys}}) \le C(\theta_{\text{dev}}),
\label{eq:cost model 2}
\end{equation}
even though the exact dollar savings remain vendor-specific. This makes moving reliability policy from fixed on-die ECC into the controller a promising path to lower HBM \$/GB without changing the HBM interface itself. 
We intentionally avoid parameterizing $p(\theta)$ with proprietary vendor data. Rather than predicting exact \$/GB, our goal is to show that once the system safely operates at $\theta_{\text{sys}} \approx 10^{-3}$, any monotone yield curve implies Eq.~\eqref{eq:cost model 2}.

\subsection{Basics of RS Codes}
\label{sec:rs-basics}

A Reed--Solomon (RS) code operates over a finite field (i.e., Galois field) $\mathrm{GF}(2^m)$, where one \textit{symbol} is $m$ bits. An $(n,k)$ \textit{systematic} RS code appends $r\triangleq n-k$ parity symbols to $k$ data symbols. Let $\mathbf{D}\in\mathrm{GF}(2^m)^k$ be the data vector and $\mathbf{P}\in\mathrm{GF}(2^m)^r$ be the parity vector. Parity is a fixed linear map
\begin{equation}
\label{eq:rs-gout}
\mathbf{P} = \mathbf{G}_{\text{out}}\mathbf{D}, \qquad
\mathbf{G}_{\text{out}}\in\mathrm{GF}(2^m)^{r\times k}.
\end{equation}
Decoding guarantees follow the standard symbol budget
\begin{equation}
\label{eq:rs-budget}
2T_{\text{sym}} + E_{\text{sym}} \le r,
\end{equation}
where $T_{\text{sym}}$ unknown-position errors and $E_{\text{sym}}$ known-position erasures are jointly correctable. Let $\{a_j\}_{j=0}^{n-1}\subset\mathrm{GF}(2^m)^\ast$ be fixed, distinct evaluation points. For a received stream $\{y_j\}_{j=0}^{n-1}$ with $t$ unknown-position errors and $e$ erasures ($2t{+}e\le r$), a conventional decoder runs four steps that map cleanly to hardware:

\begin{enumerate}
  \item \textit{Syndrome formation.}
  Stream once over the codeword and accumulate $r$ checks:
  \begin{equation}
  \label{eq:rs-syndrome}
  S_\ell = \sum_{j=0}^{n-1} y_ja_j^{\ell}, \qquad \ell=0,\ldots,r-1.
  \end{equation}
  This performs $O(nr)$ field operations per codeword with $O(r)$ state. With $O(r)$ parallel accumulators the latency is one pass (\emph{time} $O(n)$).

  \item \textit{Key-equation solve.}
  Reduce the $r$ syndromes to a small set of coefficients (e.g., an error locator and an evaluator). Berlekamp--Massey runs in $O(t^2)$ (typical in practice), while extended Euclid gives a safe $O(r^2)$ bound. The stage is narrow and serialized within a codeword. Throughput comes from replication or time-interleaving.

  \item \textit{Position locate.}
  If positions are unknown ($t>0$), evaluate the degree-$t$ locator at the $n$ candidates $\{a_j\}$. The arithmetic budget is $O(nt)$ per codeword. Using standard Chien recurrences, a fixed-width engine with $O(t)$ state produces about one candidate evaluation per cycle, so \emph{time} is $O(n)$. Vectorizing across $P$ evaluators gives $O(n/P)$ time with roughly $P$-fold datapath cost. As $n$ grows, this sweep often \textbf{dominates area and tail cycles}.

  \item \textit{Value correction.}
  With positions known (from erasures or Step~3), compute magnitudes and update $y_j$. Work is $O(t{+}e)$ and parallelizes across fixes. Inversions can be shared or avoided via composite fields or log/antilog tables.
\end{enumerate}

\begin{figure}[htbp]
  \centering
  \includegraphics[width=\linewidth]{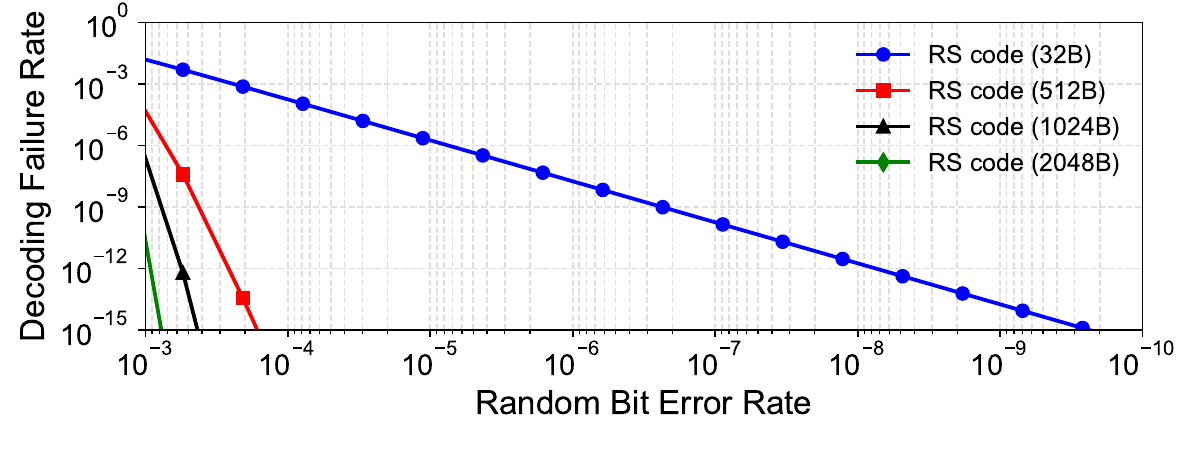}
  \caption{Decoding failure rate vs.\ codeword size at a fixed code rate (16/17). Larger codewords tolerate higher raw BER.}
  \label{fig:rs_fail_vs_n}
\end{figure}

At a fixed code rate $k/n$, increasing $n$ raises the minimum distance $d_{\min}=r{+}1$ and lowers decoding failure for the same raw BER. Fig.~\ref{fig:rs_fail_vs_n} shows this effect for a 16/17 rate: moving from tens of bytes to kilobytes per codeword reduces decoding failure by orders of magnitude for the same raw BER. 
At fixed link throughput, sustaining TB/s requires parallelizing the locate step across many evaluators, the locator datapath and candidate sweep dominate area/power as n grows.
The field size $m$ primarily affects constants (e.g., bit-parallel GF multiplier area/energy grows roughly with $m^2$). In hardware terms: Steps~1 and~4 are streaming and scale with datapath width. Step~2 is narrow but serialized within a codeword. Step~3 grows with $n$ and typically sets the long-codeword critical path. When all bad positions are provided as erasures (i.e., the unknown-error count is $t{=}0$), the position-locate stage is unnecessary and the path reduces to the streaming work above.

\subsection{Using Long RS Codes in HBM}
\label{sec:opportunity}

Section~\ref{sec:rs-basics} established two facts about RS decoding: syndromes and value correction stream and stay light, while the position-locate step scales with code length $n$ and dominates cost for long codewords. We now ask how the RS codes play out in HBM.

HBM3 exposes a hard 32\,B transaction at very high symbol rates. Vendors hold raw BER low with on-die, short-code protection, which raises test/binning pressure and \$/GB. If a system can tolerate a higher device BER but still guarantee end-to-end integrity, the cost pressure eases. With the PHY and the 32\,B packet fixed, the only lever available at the system boundary is codeword length. Longer RS codewords increase parity span and recover reliability headroom without touching the link.
Fig.~\ref{fig:rs_fail_vs_n} shows the basic effect at a fixed code rate (16/17): moving from tens of bytes to kilobytes per codeword lowers decoding failure by orders of magnitude for the same raw BER. This argues that long codes are the right reliability knob when the PHY is fixed.

Directly implementing long codes inside DRAM dies is impractical. DRAM logic budgets, tight power, and conservative clocks leave little room for wide finite-field engines or large locator arrays. The memory controller is the practical home for strong codes. It can host high-throughput datapaths, select parameters per region, and schedule work against bandwidth demand.

\begin{figure}[htbp]
    \centering
    \includegraphics[width=.9\linewidth]{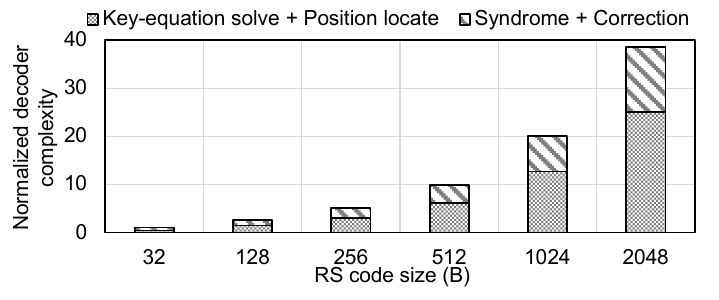}
    \caption{Normalized decoder complexity vs.\ codeword size at 1TB/s throughput (1GHz). For each size we select the minimum feasible \( \mathrm{GF}(2^m) \) to minimize total silicon. Global PHY/SerDes is excluded.}
    \label{fig:codeword_complexity}
\end{figure}

However, naively adopting long codewords in the controller introduces two immediate systems problems:

\begin{enumerate}
  \item \textit{Small-access amplification.} A single 32\,B write can trigger a full-block read or read–modify–write. 
  For a 2\,KB codeword with 128 parity, rebuilding parity from the block moves 2176B for a 32\,B update (i.e., $68\times$ amplification).

  \item \textit{Decoder silicon cost growth with $n$.} As reviewed in Section~\ref{sec:rs-basics}, the position locate step must test $n$ candidates and becomes the dominant source of area, power, and tail cycles as codewords reach kilobytes. Fig.~\ref{fig:codeword_complexity} quantifies this trend for designs provisioned to sustain a 1\,TB/s link: compared to a 32\,B RS decoder, a 2\,KB decoder is $38.6\times$ more complex, and within the 2\,KB design the locator is $1.8\times$ the check/correct logic.
\end{enumerate}

These constraints force two concrete design questions. First, how do we avoid span-scale work when only few small 32\,B units are touched. Second, how do we keep the long-code path small and fast enough to meet multi-TB/s without the locator dominating area/power and tail latency. The next section answers both by organizing reliability so 32\,B operations complete locally on the common path, and the rare long-span repair runs a short, bounded-cost path consistent with the RS decoder structure.

\section{Proposed Solution: REACH}
\label{sec:method}




To resolve the barriers of small-access amplification and decoder complexity identified in Section~\ref{sec:opportunity}, we move reliability enforcement from the device to the controller. We introduce \textbf{REACH} (\textbf{R}eliability \textbf{E}xtension \textbf{A}rchitecture for \textbf{C}ost-effective \textbf{H}BM). By decoupling system integrity from physical device characteristics, REACH enables operation at high raw BER (e.g., $10^{-3}$) while preserving the standard HBM PHY and 32\,B interface.

\begin{figure}[htbp]
    \centering
    \includegraphics[width=\linewidth]{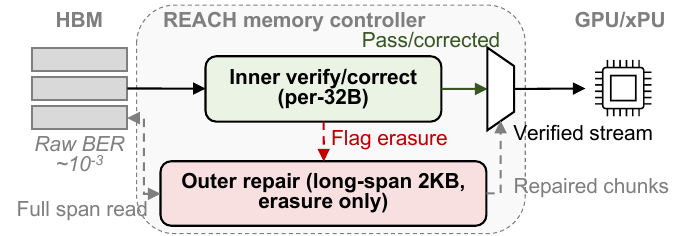}
    \caption{REACH high-level architectural overview. The design bifurcates memory access to address high device BER: (1) A \textbf{fast path} (green) filters common-case traffic locally at 32\,B granularity. (2) A \textbf{reliability path} (red) invokes the long-span RS engine in \textbf{erasure-only} mode solely when the inner code flags an erasure, eliminating locator logic overhead.}
    \label{fig:reach-intro}
\end{figure}

As illustrated in Fig.~\ref{fig:reach-intro}, REACH acts as a reliability filter that bifurcates memory traffic. A \textbf{fast path} processes the vast majority of requests using a per-32\,B inner RS code, verifying integrity or flagging explicit erasures locally. A \textbf{reliability path} serves as a safety net, invoking a long-span outer RS code only when erasures occur. Crucially, because error locations are known from the inner layer, the outer engine operates in \textit{erasure-only} mode, eliminating the expensive locator logic typically required for long codes.

Complementing this read architecture, REACH handles random writes using differential parity updates and employs an importance-adaptive bit-plane policy. The remainder of this section details how this architecture resolves the specific barriers identified in Section~\ref{sec:opportunity}: bounding amplification on small accesses (Section~\ref{sec:q1-small}), eliminating locator complexity from long-code decoding (Section~\ref{sec:q2-service}), and leveraging data semantics for further optimization (Section~\ref{sec:importance}).

\subsection{Avoiding Small-Access Amplification}
\label{sec:q1-small}
Longer ECC is necessary to tolerate higher raw error rates at fixed PHY and 32\,B transfers. Once the protection span reaches kilobytes, \textbf{it is natural to ask how a single 32\,B touch can avoid paying span-scale cost.}

\smallskip
\noindent\textbf{Naive random write under long ECC.}
A chunk is 32\,B. A long ECC span covers $W$ bytes, which equals $N{=}W/32$ chunks, and holds $P$ parity bytes. Touching one chunk with a naive handler triggers read–modify–write over the full span: read all $W$, recompute parity over the span, then write data and parity back. The I/O traffic and amplification for one 32\,B update are
\begin{equation}
T_{\text{naive}} = W + P,
\qquad
\mathrm{Amp}_{\text{naive}} = \frac{W + P}{32} = N + \frac{P}{32}.
\label{eq:naive update}
\end{equation}
For example, let the long ECC span be $W{=}2048$\,B, so $N{=}64$ chunks, and let the parity be $P{=}128$\,B for that span. Consider a \textit{random write} to one 32\,B chunk. The naive handler moves
$T_{\text{naive}}=2176\text{B}$, $\mathrm{Amp}_{\text{naive}}=68\times$.
This bound does not rely on workload assumptions. Without a way to validate the target 32\,B in place, parity must be rebuilt from the full span.

\begin{figure}[htbp]
    \centering
    \includegraphics[width=\linewidth]{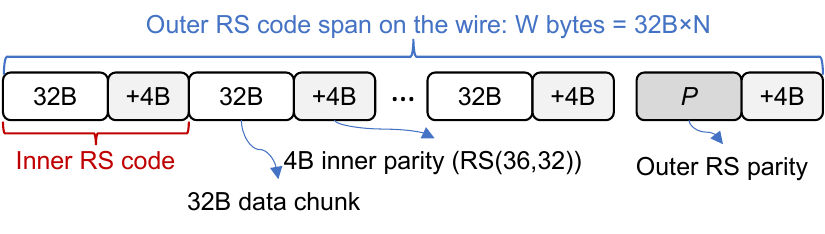}
    \caption{Illustration of REACH's two-level ECC. Each chunk carries inner RS(36,32) for local accept/correct or erasure flag. The outer RS spans $W$ bytes and maintains parity $P$. It repairs only flagged chunks as known erasures.}
    \label{fig:two-level-ECC-illustration}
\end{figure}

\smallskip
\noindent\textbf{Local decisions via REACH's two-level design.}
This necessitates a per-chunk check so the controller can decide locally and only touch the long span when required. REACH uses RS code for ECC. As shown in Fig.~\ref{fig:two-level-ECC-illustration}, REACH uses a two-level code organization:
\begin{itemize}
    \item \textit{Inner RS code (per chunk).} Each chunk is encoded by RS(36,32): 32\,B data + 4B parity. On a read, run inner RS decode. If no error is detected, accept the chunk. If an error is detected and it is within the RS(36,32) correction capability (up to 2B), correct and accept. Otherwise reject the chunk. On a write, run \textit{inner RS encode} to produce the 4B inner parity for the new 32\,B.

    \item \textit{Outer RS code (over the long span).} The $W$-byte span, which covers $N$ chunks, is encoded by an outer RS code. The controller repairs only chunk indices that the inner code rejects. The PHY and the 32\,B transfer size stay unchanged.
\end{itemize}

\smallskip
\noindent\textbf{Random write under REACH.}
A random write touches a small set of 32\,B chunks inside one outer RS codeword of size $W$ bytes. We refer to the naive bound in Eq.~\eqref{eq:naive update}, which requires moving $T_{\text{naive}}=W+P$ bytes for any single-chunk update. We now show the fast path that avoids this full-span work, followed by the bounded fallback when the inner RS rejects a chunk.

\emph{Fast path via differential parity when inner RS accepts.}
If the inner RS accepts or locally corrects all touched chunks, the controller updates only those chunks and the outer parity. We refer to Eq.~\eqref{eq:rs-gout}, $\mathbf{P}=\mathbf{G}_{\text{out}}\mathbf{D}$ over $\mathrm{GF}(2^{16})$. Construct two sparse data vectors $\mathbf{D}_{\text{old}}$ and $\mathbf{D}_{\text{new}}$ by placing the old and new 32\,B payloads of the $q$ updated chunks at their original symbol positions and zeros elsewhere. Linearity gives a local parity update:
\begin{equation}
\label{eq:diff-parity-sparse}
\mathbf{P}_{\text{new}}
=
\mathbf{P}_{\text{old}}
\oplus
\mathrm{RS}(\mathbf{D}_{\text{new}})
\oplus
\mathrm{RS}(\mathbf{D}_{\text{old}}).
\end{equation}
This touches only the $q$ updated chunks and the $P$ parity bytes. It does not depend on the span $W$.

The fast-path traffic reads and writes each touched chunk once (36B per chunk on the wire) and writes parity once. The total and the I/O amplification relative to useful bytes $32q$ are
\begin{equation}
\label{eq:tfast}
T_{\text{fast}}(q) = 72q + P,
\end{equation}
\begin{equation}
\label{eq:ampfast}
\mathrm{Amp}_{\text{fast}}(q)
=
\frac{T_{\text{fast}}(q)}{32q}
=
2.25 + \frac{P}{32q}.
\end{equation}
For instance, with $W=2048$\,B and $P=128$\,B, Eq.~\eqref{eq:ampfast} gives $6.25\times$, $4.25\times$, and $3.25\times$ for $q\in\{1,2,4\}$, compared to the naive $68\times$ from Eq.~\eqref{eq:naive update}. This improvement follows directly from linearity and does not rely on workload assumptions.

\smallskip
\emph{Outer RS repair when inner RS rejects.}
If the inner RS rejects any touched chunk, the controller runs one erasure-only outer repair on the flagged indices, applies the payload update, recomputes parity once via Eq.~\eqref{eq:diff-parity-sparse}, and commits in data-before-parity order. As $r$ defined in Eq.~\eqref{eq:rs-budget}, then with chunk-level capacity $C$ from
\begin{equation}
\label{eq:chunk-capacity}
E_{\text{chunk}} \le \Big\lfloor \frac{r}{16} \Big\rfloor \triangleq C,
\end{equation}
when the number of rejected chunks $E$ satisfies $E\le C$, the bounded repair traffic is
\begin{equation}
\label{eq:trepair}
T_{\text{repair}} \le W + P,
\end{equation}
and no second repair pass is needed. If all touched chunks pass in inner RS, the outer code is not invoked.

\begin{figure}[htbp]
    \centering
    \includegraphics[width=\linewidth]{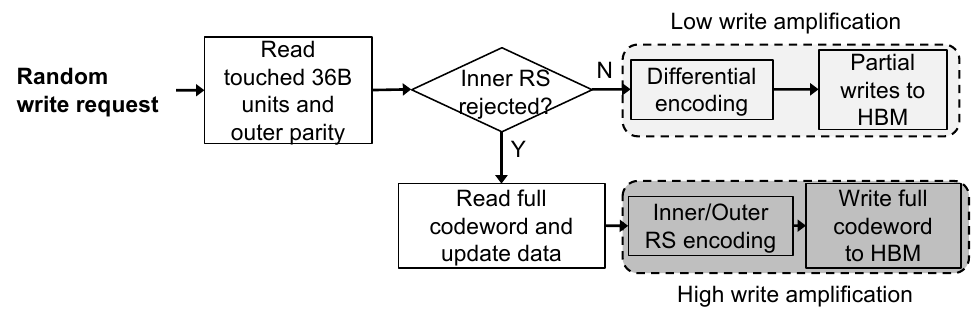}
    \caption{Random write flow under REACH.}
    \label{fig:randwrite_flow}
\end{figure}

Fig.~\ref{fig:randwrite_flow} summarizes the controller actions. The controller reads the $q$ touched chunks and the outer parity, runs inner RS on each 32\,B unit, and decides locally. If all touched chunks are accepted or corrected, it computes the outer parity delta for those columns via Eq.~\eqref{eq:diff-parity-sparse}, writes $q\times36$B of data and $P$ of parity to HBM. The resulting traffic follows Eq.~\eqref{eq:tfast}–\eqref{eq:ampfast}. If any touched chunk is rejected, the controller escalates once. It repairs the flagged positions with the outer RS codeword, applies the payload update, recomputes parity once, and commits data before parity. The traffic is bounded by Eq.~\eqref{eq:trepair}, and there is no second repair pass. This two-step flow makes the common case local and low-amplification.

\begin{figure}[htbp]
    \centering
    \includegraphics[width=\linewidth]{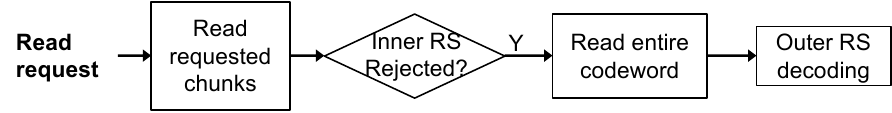}
    \caption{Random-read flow under REACH.}
    \label{fig:randread_flow}
\end{figure}

\smallskip
\noindent\textbf{Random read under REACH.}
A read request asks for $q$ 32\,B chunks inside one outer RS codeword of size $W$ with $P$ parity bytes. Fig.~\ref{fig:randread_flow} shows the flow. The controller first reads only the requested chunks and runs inner RS on each 32\,B unit. If all touched chunks are accepted or locally corrected, it returns the bytes and completes without invoking the outer code.

If any touched chunk is rejected, the controller escalates once. It reads the full outer RS codeword (data plus parity), repairs the flagged positions, and then returns the requested bytes. There is no second repair pass. This per-chunk decision keeps the common case local and touches the outer code only when necessary. Without this boundary, a small read under a long code would require a span-scale fetch, as in Eq.~\eqref{eq:naive update}.

\subsection{Keeping the Controller Small and Fast}
\label{sec:q2-service}

Long-code decoders that must \textit{locate} unknown error positions dominate area and power once codewords reach kilobytes (Fig.~\ref{fig:codeword_complexity}). \textbf{We are led to consider removing the locator by deciding at 32\,B granularity.} The inner RS labels each 32\,B chunk as either \textit{accepted} or \textit{erasure}. With positions known, the outer RS performs a single erasure-only repair over exactly those coordinates and stops.

The outer RS operates over $\mathrm{GF}(2^{16})$ where one symbol is two bytes. A codeword contains $k$ data symbols and $r$ parity symbols. If an outer codeword spans $W$ bytes on the wire, the number of 32\,B chunks is:
\begin{equation}
\label{eq:n-chunks}
N = \frac{W}{32}.
\end{equation}
With $r$ parity symbols and $16$ symbols per 32\,B chunk, the erasure capacity measured in 32\,B chunks is:
\begin{equation}
\label{eq:cap-chunks}
C = \Big\lfloor \frac{r}{16} \Big\rfloor.
\end{equation}
The architecture guarantees correction as long as the number of flagged chunks $E \le C$.

\begin{figure}[htbp]
    \centering
    \includegraphics[width=\linewidth]{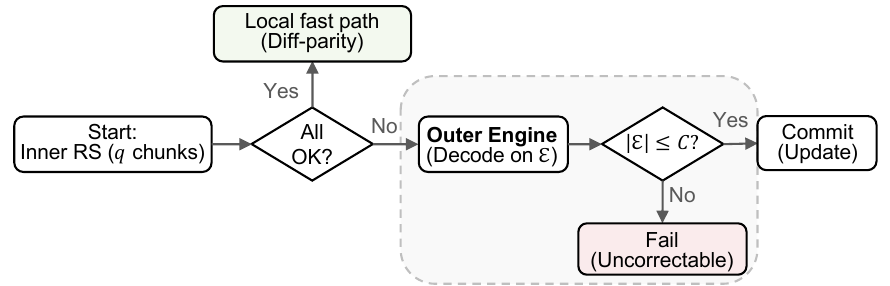}
    \caption{Path split for random read or write at the erasure boundary. The inner RS either accepts or corrects each touched 32\,B chunk.}
    \label{fig:erasure-decode}
\end{figure}

\smallskip
\noindent\textbf{Erasure-only execution flow.}
This design reduces the complex mathematical problem of error correction into a deterministic scheduling task. As shown in Fig.~\ref{fig:erasure-decode}, the controller operates in two phases:
\begin{enumerate}
    \item \textbf{Accumulate:} The Inner RS checks each 32\,B chunk. Clean chunks pass immediately to the system. Failed chunks are logged into an \textit{erasure set} $\mathcal{E}$.
    \item \textbf{Repair:} If $\mathcal{E} \neq \varnothing$, the controller stalls the reliability path, retrieves the outer parity, and triggers the Outer Erasure Engine to solve for the missing chunks in $\mathcal{E}$.
\end{enumerate}

Crucially, this eliminates the iterative search logic (e.g., Chien search) from the decoder. The complexity of repair becomes a deterministic function of the erasure count $|\mathcal{E}|$ rather than the codeword length $N$. We provide a rigorous verification of this capacity ($C$) against high raw BER and correlated burst faults in Section~\ref{sec:verification}.

\subsection{Importance-Adaptive Bit-Plane ECC}
\label{sec:importance}
With small-access amplification removed and the outer path simplified by erasures, we are still left asking a practical question: \textbf{can decoder silicon go down further at multi-TB/s?} The key observation is that not all bits contribute equally to inference quality.

\begin{figure}[htbp]
  \centering
  \includegraphics[width=\linewidth]{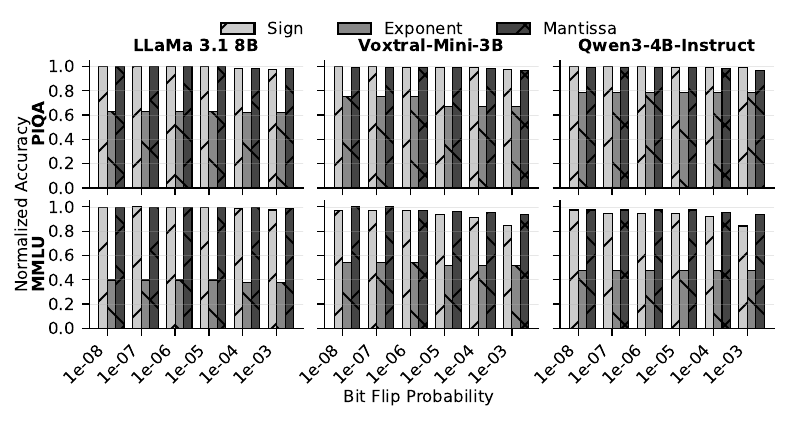}
  \caption{Motivational microbenchmark: the critical vulnerability of exponent bits in unprotected BF16 models. Accuracy is evaluated on PIQA (top) and MMLU (bottom) after injecting bit-flips into the sign, exponent, and mantissa.}
  \label{fig:bitfield_sensitivity}
\end{figure}

Fig.~\ref{fig:bitfield_sensitivity} shows a simple microbenchmark on BF16 LLaMA-3.1-8B~\cite{llama-3-1-8b-instruct}, Voxtral-Mini-3B~\cite{Voxtral-Mini-3B-2507}, Qwen3-4B~\cite{qwen3-4b-instruct-2507}. Injecting random bit-flips into the exponent field rapidly destroys PIQA~\cite{bisk2020piqa} and MMLU~\cite{hendrycks2020measuring} accuracy, while the same fault rate in mantissa bits causes only mild degradation. In other words, exponent bits are fragile, whereas most mantissa bits are far less so.

\begin{figure}[htbp]
    \centering
    \includegraphics[width=\linewidth]{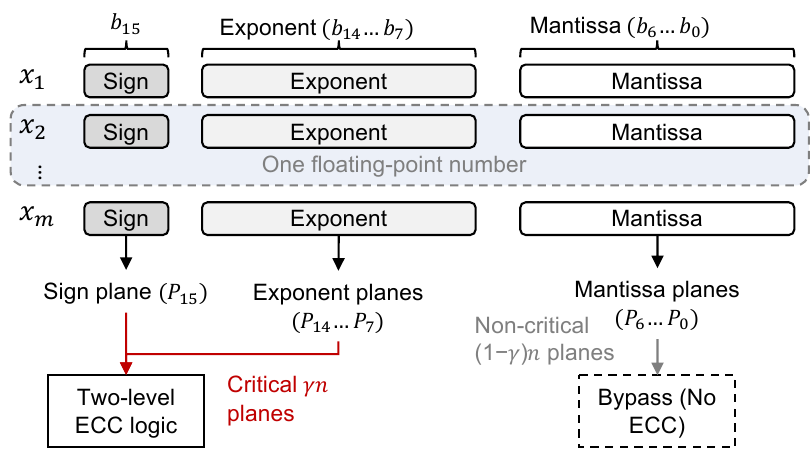}
    \caption{Illustration of the importance-adaptive ECC protection. A block of $m$ BF16 numbers is stored in a bit-plane-oriented layout. The critical bit-planes are extracted and processed by the ECC logic. The non-critical planes bypass the ECC logic, reducing controller overhead and write traffic.}
    \label{fig:bit-plane-layout}
\end{figure}

This gap creates room for importance-adaptive protection. 
To support this functionality efficiently, we adopt a bit-plane–oriented in-memory data placement scheme as shown in Fig.~\ref{fig:bit-plane-layout}.
Let a block contain $m$ numeric values $\{x_1,\dots,x_m\}$. Each value has $n$ bits, written as
$x_j=[b_{j,n-1},\dots,b_{j,0}]$ where $b_{j,i}\in\{0,1\}$ is bit $i$ of $x_j$.
Define the bit-plane at position $i$ as $P_i=\{b_{1,i},\dots,b_{m,i}\}$.
Let $\mathcal{S}\subseteq\{0,\dots,n-1\}$ be the set of \textit{critical} planes that we decide to protect end to end, and let
$\gamma = \frac{|\mathcal{S}|}{n}$
be the fraction of planes that are protected. In this work, we treat $\gamma$ as a \textit{model-level} tuning knob rather than a per-layer parameter. 
Planes in $\mathcal{S}$ use the two-level ECC: inner RS at 32\,B granularity and outer RS across the long span.
Planes not in $\mathcal{S}$ bypass the \textit{outer} RS and may optionally bypass the \textit{inner} per-chunk RS as a speed and area optimization.
The wire protocol and the 32\,B transfer size do not change.

Only data assembled from planes in $\mathcal{S}$ enter the outer RS codeword. As the protected-plane ratio $\gamma$ decreases, three effects follow:
(i) \textit{Outer RS work scales down.} Less protected data per long codeword means less to encode, check, and repair. The outer RS engine can be provisioned smaller, and its activity drops roughly in proportion to the protected payload.
(ii) \textit{Inner per-chunk work can also drop (optional).} If planes not in $\mathcal{S}$ skip the per-chunk inner RS, fewer 32\,B units run inner encode/decode. This reduces lane provisioning and shortens the active path on those units.
(iii) \textit{Parity storage and movement shrink.} With less protected data, the controller maintains and updates less ECC parity and moves less ECC traffic on updates.

All accesses still follow the same rule as Section~\ref{sec:q1-small} and Section~\ref{sec:q2-service}.
First, check each 32\,B unit that belongs to planes in $\mathcal{S}$ with inner RS.
Second, escalate once to the outer RS only if any such unit is rejected and only for the flagged indices.
Unprotected planes never trigger outer repair, so they do not contend for the outer decoder. With the complete architectural mechanisms defined, we next analyze their theoretical reliability bounds before evaluating system performance.

\section{Reliability Verification}
\label{sec:verification}

Before evaluating end-to-end performance, we must mathematically verify that our ``erasure-only'' outer code strategy is sufficient to bridge the gap between high raw device BER and system reliability targets. We show the closed-form example at $\mathrm{BER}=10^{-4}$ for readability. Qualification up to $\mathrm{BER}=10^{-3}$ is validated in Section~\ref{sec:evaluation}. Specifically, we must demonstrate two properties: (1) \textbf{Safety}, ensuring that the probability of exceeding the outer code's correction capability is negligible; and (2) \textbf{Performance}, ensuring that the outer decoder is invoked rarely enough to be treated as a cold path.

We base this analysis on the configuration defined in Section~\ref{sec:method}: a 2\,KB outer span ($W{=}2048$\,B) comprising $N{=}64$ data chunks, protected by 128B of outer parity ($C{=}4$ chunk-erasure capacity). We assume a conservative raw $\mathrm{BER}=10^{-4}$ with independent bit flips.

\subsection{Erasure Escalation Probability}
\label{sec:math-bounds}

First, we quantify how often the inner RS(36,32) code rejects a chunk. Let $q$ be the probability of a byte error at $\mathrm{BER}=10^{-4}$:
\begin{equation}
q = 1 - (1 - 10^{-4})^{8} \approx 8.0\times 10^{-4}.
\end{equation}
The number of faulty bytes $X$ in one 36B inner block follows a binomial distribution $X \sim \mathrm{Binomial}(36,q)$. Since the inner code corrects up to 2 bytes, it escalates only when $X \ge 3$. The per-chunk rejection probability $p_{\text{rej}}$ is:
\begin{equation}
p_{\text{rej}} = P(X \ge 3) = 1 - \sum_{k=0}^{2} \binom{36}{k} q^k (1-q)^{36-k} \approx 3.6\times 10^{-6}.
\end{equation}
Intuitively, while $\sim$2.8\% of chunks contain at least one bit flip, 99.99\% of those errors are corrected locally. Only one in $\sim$270,000 chunks is ``hard'' enough to trigger an erasure.

Next, we verify the stability of the outer code. For a 2\,KB codeword containing $N{=}64$ data chunks, the expected number of erasures is $\mu \triangleq N \cdot p_{\text{rej}} \approx 2.3 \times 10^{-4}$. The outer code fails only if the number of erasures $E$ exceeds the capacity $C=4$. We use the Poisson tail envelope to bound this failure probability:
\begin{equation}
\label{eq:poisson-tail}
P(E > C) \le \sum_{j=C+1}^{\infty} \frac{\mu^{j}}{j!}e^{-\mu}
\le \frac{\mu^{C+1}}{(C{+}1)!}e^{-\mu}.
\end{equation}
Plugging in our parameters ($\mu \approx 2.3 \times 10^{-4}, C=4$):
\begin{equation}
P(E > 4) \lesssim \frac{(2.3 \times 10^{-4})^5}{120} < 10^{-18}.
\end{equation}
Table~\ref{tab:repair-breakdown} summarizes the probabilities at each layer. The result confirms the safety condition: the probability of an uncorrectable error per codeword is effectively zero, far exceeding the reliability of standard on-die ECC.

\begin{table}[htbp]
    \centering
    \caption{Hierarchical repair probability at BER=$10^{-4}$. Most errors are fixed locally; outer repair is rare; failure is negligible.}
    \label{tab:repair-breakdown}
    \begin{tabular}{l l l}
        \toprule
        \textbf{Layer} & \textbf{Outcome} & \textbf{Probability} \\
        \midrule
        \textbf{Inner RS} & Clean ($X=0$) & $0.9716$ \\
        (Per 32\,B) & Local Fix ($X \in \{1,2\}$) & $2.84\times 10^{-2}$ \\
         & \textbf{Escalate ($X \ge 3$)} & $\mathbf{3.6\times 10^{-6}}$ \\
        \midrule
        \textbf{Outer RS} & No Erasures ($E=0$) & $0.99977$ \\
        (Per 2\,KB) & Repaired ($1 \le E \le 4$) & $2.3\times 10^{-4}$ \\
         & \textbf{Uncorrectable ($E > 4$)} & $\mathbf{< 10^{-18}}$ \\
        \bottomrule
    \end{tabular}
\end{table}

\smallskip
\noindent\textbf{Validity under burst faults.}
While the analytical model above assumes independent bit flips, REACH remains architecturally robust to correlated bursts (e.g., TSV failures) common in real devices. The Inner RS acts as a \textit{fault normalizer}: it collapses any error pattern within a 32\,B unit, whether a sparse 3-bit flip or a dense 256-bit burst, into a single binary outcome (i.e., an erasure). Consequently, the safety of the system depends only on the \textit{count} of faulty chunks, not their internal error distribution. 
Robustness depends on whether correlated faults consume the outer code’s erasure budget within a codeword. If telemetry shows higher multi-chunk erasure rates, the system can increase parity $r$, shorten span $W$, or use interleaving to spread correlated faults across codewords.
Given this property, we maintain the independent error model in our evaluation (Section~\ref{sec:evaluation}) as a valid stress test for the system's erasure recovery capacity, acknowledging that precise maps of physical defects rely on non-public vendor data.

\subsection{Workload-Aware Repair Frequency}
\label{sec:workload-freq}

We now map these probabilities to memory requests to validate the performance condition. We assume an LLM-inference access mix: 95\% reads / 5\% writes, and 95\% sequential / 5\% random access. 

The probability that a memory request triggers the outer decoder, $p_{\text{esc}}$, depends on how many chunks it touches.
\begin{itemize}
    \item \textbf{Sequential Read (SR):} Reads a full 2\,KB codeword ($N{=}64$ chunks).
    $p_{\text{esc}}^{\text{SR}} = 1 - (1 - p_{\text{rej}})^{64} \approx 2.3\times 10^{-4}$.
    \item \textbf{Random Read (RR):} Reads a 32\,B target. However, our architecture may speculatively fetch neighbors. Conservatively assuming a window of $m{=}32$ chunks:
    $p_{\text{esc}}^{\text{RR}} \approx 1 - (1 - p_{\text{rej}})^{32} \approx 1.1\times 10^{-4}$.
    \item \textbf{Random Write (RW):} Uses differential parity. It reads the target payload ($m$ chunks) plus the outer parity ($r_{\text{pos}}{=}4$ chunks).
    $p_{\text{esc}}^{\text{RW}} = 1 - (1 - p_{\text{rej}})^{m + r_{\text{pos}}} \approx 1.3\times 10^{-4}$.
\end{itemize}

Weighting these by the access mix ($p_{\text{SR}} \approx 0.90$, $p_{\text{RR}} \approx 0.05$, $p_{\text{RW}} \approx 0.05$):
\begin{equation}
p_{\text{outer}} \approx \sum_{type} p_{type} \cdot p_{\text{esc}}^{type} \approx 2.1\times 10^{-4}.
\end{equation}
This establishes the performance result: \textbf{roughly only one in every 5,000 requests invokes the outer decoder.} The remaining 99.98\% of requests are handled entirely by the fast inner path. This confirms that the heavy outer decoder acts as a cold standby unit rather than a throughput bottleneck, which justifies the area-efficient ``erasure-only'' design.


\begin{figure*}[htbp]
  \centering
  \includegraphics[width=\linewidth]{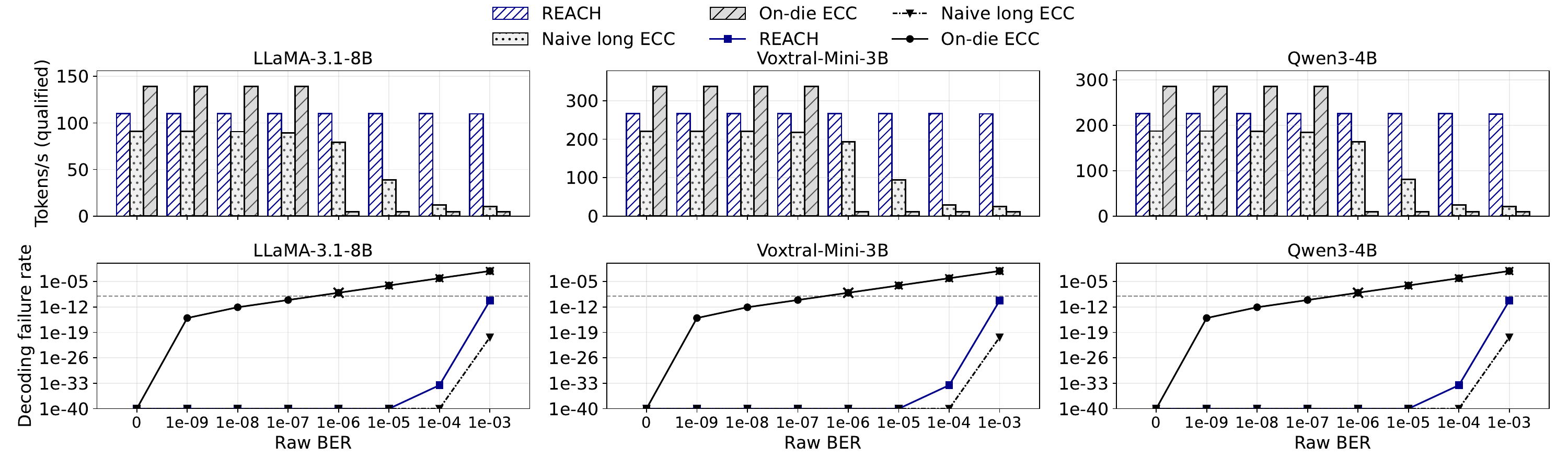}
  \caption{Qualified tokens/s (top, higher is better) and per-token decoding failure rate (bottom, lower is better) vs.\ raw BER at 8K context for LLaMA-3.1-8B, Voxtral-Mini-3B, and Qwen3-4B. Configuration: \textbf{REACH} (inner RS(36,32) + fixed 2\,KB outer RS code), \textbf{naive long ECC} (a 2\,KB RS code with naive RMW), and \textbf{on-die ECC}. A point is qualified when the per-token failure rate $\le 10^{-9}$ (bottom, gray dashed line).}
  \label{fig:e2e}
\end{figure*}

\section{Evaluation}\label{sec:evaluation}

\subsection{Experimental Setup}
\label{sec:setup}

We evaluate REACH using a trace-driven simulation framework that couples compute scheduling with detailed memory timing.
We first profile each model in SCALE-Sim~\cite{raj2025scale}, running the published systolic array in USER mode. This configuration enforces the target HBM bandwidth and emits per-layer cycle counts and time-stamped DRAM traces that explicitly account for bandwidth stalls.
Random-access ratios (4\% for LLaMA-3.1-8B, 3\% for Voxtral-Mini-3B, 4\% for Qwen3-4B) are derived directly from these SCALE-Sim traces.
The traces are then replayed inside a modified DRAMSim3~\cite{li2020dramsim3} configured for HBM timing and augmented with REACH's two-level ECC logic.

\smallskip
\noindent\textbf{Fault injection methodology.}
We run end-to-end Monte Carlo fault injection at 8K context, sweeping raw BER in \{0, $10^{-9}$, $10^{-8}$, \dots, $10^{-3}$\}. For every BER point, we flip each memory bit independently. DRAMSim3 generates the resulting per-request service times—including latencies for outer-code repairs or replays—which are folded back onto the SCALE-Sim schedule to compute the full decode time.

\smallskip
\noindent\textbf{Configurations.}
We compare three distinct reliability architectures:
(1) \textit{REACH}: Uses an inner RS(36,32) on each 32\,B payload chunk and a fixed 2\,KB outer RS codeword with a $\sim$0.9 code rate. The outer RS decoder array is provisioned to run well below saturation.
(2) \textit{Naive long ECC}: Implements a 2\,KB RS code that performs a naive read–modify–write across the whole codeword for small updates.
(3) \textit{On-die ECC}: Represents the standard HBM baseline.

\smallskip
\noindent\textbf{Calibration and metrics.}
We report the per-token decoding failure rate and ``qualified tokens/s.'' The latter is defined as the inverse of the total runtime (updated with DRAMSim3 delays) when the failure rate is $\le 10^{-9}$, and 0 otherwise. We calibrate the BER=0 point to an NVIDIA H100-SXM (80GB HBM3, 3.35\,TB/s) by matching the compute-bound SCALE-Sim throughput, then applying code-rate and decoder-overhead factors to obtain effective coded throughput.

\subsection{End-to-End Performance and Reliability Analysis}
\label{sec:e2e-ecc}

Fig.~\ref{fig:e2e} summarizes the qualified throughput and decoding failure rates across the BER sweep. At BER=0, on-die ECC delivers 139.3 tokens/s for LLaMA-3.1-8B. REACH (configured with a 2\,KB outer code) achieves 110.1 tokens/s ($\sim$79\% of on-die), while naive long ECC reaches 90.8 tokens/s ($\sim$65\% of on-die). Voxtral-Mini-3B and Qwen3-4B show similar ratios: REACH retains about 80\% of on-die throughput, whereas naive long ECC sits in the mid-60\% range. This one-time drop at BER=0 is expected: on-die ECC keeps parity on die and pays no off-die bandwidth, whereas both RS-based designs transmit parity and pay the code-rate tax.

On-die ECC stays qualified only up to raw BER $\approx 10^{-7}$ and becomes unqualified at $10^{-6}$ and above, so its qualified tokens/s drops to zero past that point. In contrast, REACH remains qualified up to raw BER $\sim 10^{-3}$ and keeps throughput nearly flat. For LLaMA-3.1-8B, tokens/s is 110.1 at BER=$10^{-5}$ and 109.8 at BER=$10^{-3}$. The other two models exhibit similarly flat curves. Naive long ECC is qualified across the sweep but yields lower throughput at every BER point.

The gap is most visible at high BER. At BER=$10^{-3}$, the REACH delivers roughly 11$\times$ more qualified tokens/s than naive long ECC for LLaMA-3.1-8B (similar 10–11$\times$ gains for Voxtral-Mini-3B and Qwen3-4B). Even at BER=$10^{-5}$, REACH is still about $2.8\times$ faster across all three models. The reason is simple: the naive long RS design pays full-codeword read–modify–write as errors increase, while REACH fixes most faults locally and avoids span-scale RMW on the hot path. Overall, REACH extends tolerable raw BER by about three orders of magnitude compared to on-die ECC (qualified up to $\sim 10^{-3}$ vs.\ $\sim 10^{-6}$) while keeping throughput close to the BER=0 calibration, at the cost of a single $\sim$21\% hit at error-free operation.

\subsection{Sensitivity to Access Patterns}
\label{sec:access-patterns}
The end-to-end results in Fig.~\ref{fig:e2e} average over whatever mix of sequential and random accesses the LLM trace happens to generate. This shows that REACH organization can sustain high tokens/s at a given BER, but not how sensitive it is to the access pattern itself. LLM decoders are dominated by sequential streams, yet real systems always see a tail of small, scattered reads and writes from KV management, metadata, and multi-tenant interference. Once the code parameters and PHY width are fixed, the only remaining system lever is the spatial locality of requests. If performance collapses as soon as the pattern departs from ideal streaming, higher-level scheduling or caching cannot repair it. We therefore sweep the random-access and write ratios explicitly to measure how much headroom the controller actually has.

\subsubsection{Random Access}
\label{sec:random}

\begin{figure}[htbp]
\centering
\includegraphics[width=.95\linewidth]{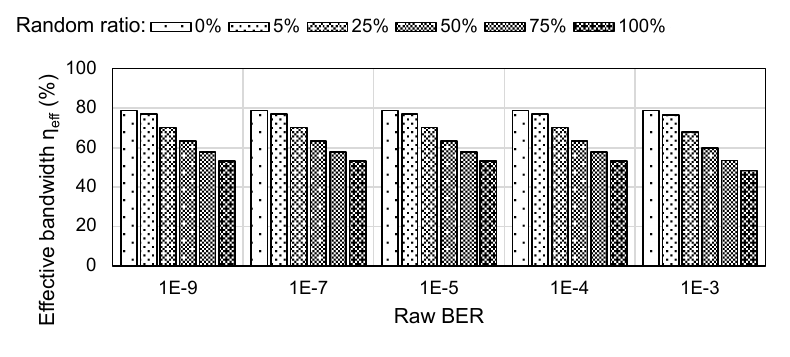}
\caption{Effective bandwidth versus random-access ratio (0\%–100\%) with a 2\,KB outer RS codeword and a fixed 5\% write share.}
\label{fig:random ratio}
\end{figure}

We define \textit{effective bandwidth} as the ratio of useful payload bytes delivered to the host over the total bytes moved on the HBM bus (data, ECC/parity, and any outer-RS escalations):
$\eta_{\text{eff}} = \frac{\text{useful bytes}}{\text{total bus bytes}}$.
Fig.~\ref{fig:random ratio} sweeps the random-access ratio from 0\% to 100\% with a fixed 5\% write share and a 2\,KB outer RS codeword in REACH. At 0\% random, $\eta_{\text{eff}}$ is flat at 78.8\% across all BER values: transfers stream full 2\,KB codewords, and the only loss is redundancy. This ceiling is dictated by the composite code rate: the outer RS uses 64 data chunks and 8 parity chunks (8/9 rate), the inner RS uses 32~B data + 4~B parity per chunk (32/36 rate), so the best case payload share is $(8/9)\times(32/36)\approx 0.79$. The small gap from 79\% to 78.8\% comes from the fixed 4\% write share and a mild 5\% write penalty in the traffic model.

As randomness increases, more requests touch only a few 32\,B chunks per codeword, so any inner RS failures trigger outer RS repair on a full codeword and inflate traffic. At 5\% random, effective bandwidth drops only slightly from 77.0\% to 76.4\% as BER rises to $10^{-3}$ (0.6 percentage points, p.p.). At 25\% random, it falls from 70.3\% to 68.1\% (2.2 p.p.); at 50\% random, from 63.5\% to 59.9\% (3.6 p.p.); at 75\%, from 57.8\% to 53.5\% (4.3 p.p.); and at 100\%, from 53.1\% to 48.3\% (4.8 p.p.). Even when every access is random and the device runs at BER=$10^{-3}$, the controller still delivers just under half of the raw link bandwidth in useful payload. Once we fix long codewords and PHY width, this is exactly the regime where random-access amplification would surface. The fact that the trend decay smoothly rather than collapsing indicates that the 32\,B inner decision boundary is doing its job.

\begin{figure}[htbp]
\centering
\includegraphics[width=\linewidth]{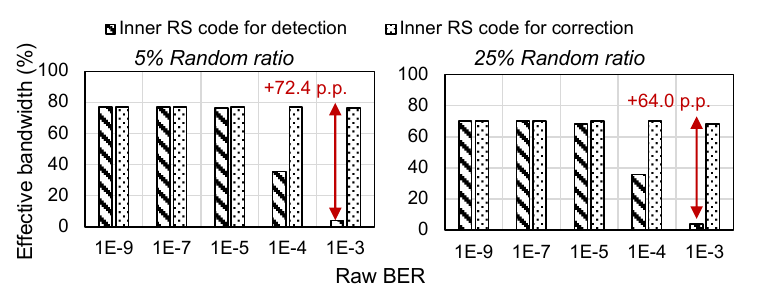}
\caption{Impact of inner RS policy on effective bandwidth under randomness (outer RS codeword = 2048B, write ratio = 5\%).}
\label{fig:inner-rs-randomness}
\end{figure}

Inner RS correction contains errors at the chunk level and limits outer RS activations, which preserves effective bandwidth even when accesses are random. The inner RS code fixes most chunks on the spot and flags only the few bad chunks as erasures, which avoids pulling an entire outer RS codeword on most random touches. We quantify this in Fig.~\ref{fig:inner-rs-randomness} by switching the inner tier to \textit{detection only} and re-measuring effective bandwidth. At BER=$10^{-3}$ with 5\% random, effective bandwidth jumps from 4.04\% (detection only) to 76.4\% (with correction). At 25\% random, it rises from 4.04\% to 68.1\%.


\subsubsection{Read/Write Ratio}
\label{sec:rw-ratio}
We now ask how sensitive the scheme is to the fraction of writes themselves. We define the \textit{write ratio} $\rho_{\mathrm{W}}$ as the fraction of memory requests that are writes (sequential + random) out of all HBM requests. In this experiment we fix the random-access share at 5\% of requests and sweep $\rho_{\mathrm{W}}$ from 0\% (all reads) to 100\% (all writes) with the same 2\,KB outer RS codeword.

\begin{figure}[htbp]
    \centering
    \includegraphics[width=.95\linewidth]{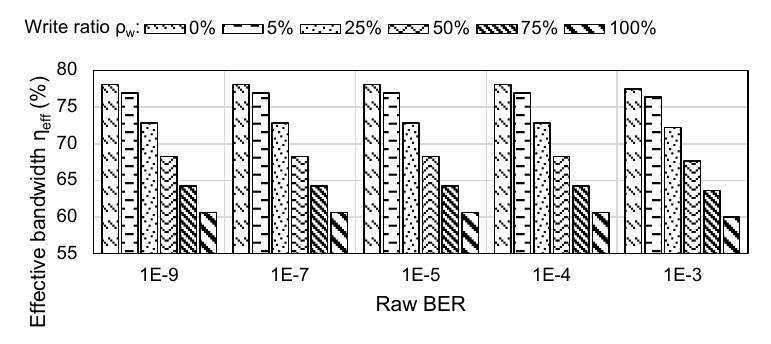}
    \caption{Effective bandwidth versus write ratio $\rho_{\mathrm{W}}$ (0–100\%) at a fixed 5\% random access share with a 2\,KB outer RS codeword.}
    \label{fig:write ratio}
\end{figure}

Fig.~\ref{fig:write ratio} reports effective bandwidth versus $\rho_{\mathrm{W}}$. At low BER ($0$–$10^{-5}$), the curves are almost perfectly linear: $\eta_{\text{eff}}$ falls from about 78\% in a read-mostly regime ($\rho_{\mathrm{W}}\approx 0$) to about 61\% when every request is a write ($\rho_{\mathrm{W}}\approx 1$). This behavior is dominated by first-order parity overhead on writes. Decoder stalls do not yet matter.
At high BER ($10^{-3}$), the entire bars shifts down by less than 1 p.p. at every write ratio, because the inner RS still corrects locally and keeps outer RS decoding rare. All points remain qualified under the $10^{-9}$ per-token failure target. Once the code rate and outer span are fixed, the write ratio sets the primary utilization loss via additional parity bytes, and the inner RS correction can keep high-BER penalties to well under a percentage point across the whole read/write spectrum.

\subsection{Outer RS Codeword Length Sensitivity}
\label{sec:outer-length}
\begin{figure}[htbp]
    \centering
    \includegraphics[width=.9\linewidth]{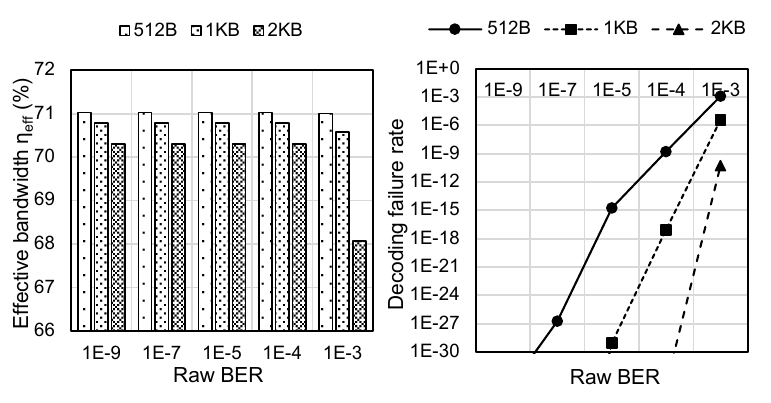}
    \caption{Impact of outer RS codeword length on effective bandwidth and decoding failure rate Left: effective bandwidth $\eta_{\text{eff}}$ versus raw BER for 512\,B, 1KB, and 2\,KB outer RS codewords at a fixed outer rate of 0.9 with inner RS(36,32), 5\% writes, 5\% random. Right: corresponding per-codeword decoding failure rate.}
    \label{fig:outer-length}
\end{figure}

We next vary the outer RS codeword length while keeping code rate fixed at 0.9 and the inner RS(36,32) unchanged. Fig.~\ref{fig:outer-length} uses a mix workload (5\% writes, 5\% random) and sweeps raw BER for 512\,B, 1KB, and 2\,KB outer codewords. On the left, effective bandwidth $\eta_{\text{eff}}$ remains tightly clustered between 71.0\% and 68.1\% across all three lengths at BER=$10^{-3}$. Moving from 512\,B to 2\,KB changes $\eta_{\text{eff}}$ by at most 2.3 p.p. On the right, the decoding failure rate drops sharply as the span grows: the 512\,B codeword remains qualified up to roughly BER=$10^{-5}$, while the 1KB and 2\,KB variants extend qualification to about $10^{-4}$ and $10^{-3}$, respectively. The tradeoff is longer outer codewords slightly perturb steady-state bandwidth but buy orders-of-magnitude more raw-BER headroom, making a 2\,KB span a natural operating point.

\subsection{Microarchitecture and PPA}
\label{sec:ppa}

We now quantify the cost of REACH's two-level ECC controller at HBM-class throughput. We implement the complete REACH controller logic in SystemVerilog and synthesize it with an open-source flow (Yosys~\cite{yosys} + OpenROAD~\cite{openroad}) using the ASAP7 7nm PDK~\cite{clark2016asap}.
The design targets BER $=10^{-3}$, a 3.35\,TB/s HBM interface, and a 20\% outer-decoder utilization budget. The synthesized netlist runs at 1.74GHz with 64 lanes of 32\,B per cycle, sustaining 3.56\,TB/s on the ECC front end.

\begin{figure}[htbp]
    \centering
    \includegraphics[width=\linewidth]{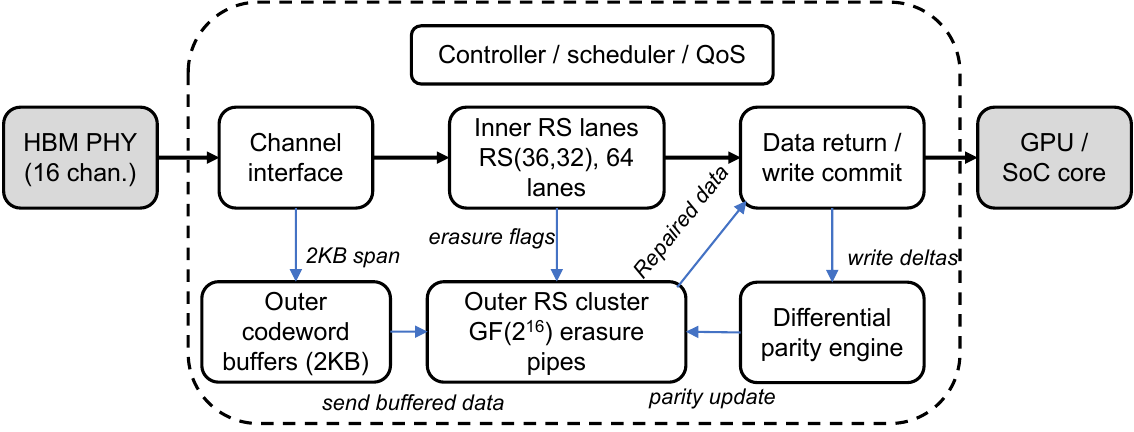}
    \caption{Block-level microarchitecture of REACH controller.}
    \label{fig:decoder-microarch}
\end{figure}

Fig.~\ref{fig:decoder-microarch} shows the block level organization. The HBM PHY drives per channel interface blocks, which strip and insert ECC metadata. These chunks feed 64 inner RS(36,32) lanes over GF$(2^8)$, one chunk per cycle per lane. Each lane decides accept, local correct, or erasure. Chunks flagged as erasures are written into 2\,KB outer codewords in a shared SRAM pool and are later repaired by a small GF$(2^{16})$ erasure cluster. Random writes go through a differential parity engine that updates only the touched outer parity symbols using the same field tables as the outer cluster.

\vspace{-10px}

At a high level, the controller instantiates 64 inner RS(36,32) lanes over GF$(2^8)$, a shared outer GF$(2^{16})$ erasure cluster, a differential parity engine for random writes, and a small buffer pool. Each inner lane processes one 32\,B chunk per cycle and decides accept, local correct, or erasure. Rejected chunks are steered to the outer cluster, which holds 26 erasure-only pipes sized for the measured $p_{\text{outer}}$ and a 20\% utilization target at 3.35\,TB/s. Random writes use differential parity to update outer parity symbols without touching the full span. The controller maintains double-buffered 2\,KB outer codewords per lane in about 320KB of SRAM, so one codeword can be repaired while the next is filled from HBM.

\smallskip
\noindent\textbf{Area and storage overhead.}
Using the synthesized gate counts, ASAP7 standard-cell areas, and the SRAM macro bit area, the complete ECC controller occupies 15.2\,mm$^2$ in ASAP7. This corresponds to about $1.7\times 10^8$ gate equivalents plus roughly 320KB of SRAM. 

\smallskip
\noindent\textbf{Energy and power overhead.}
From the synthesized netlist and ASAP7 pin capacitances, the controller consumes 17.5\,W at 1.74\,GHz and 3.56\,TB/s ($\sim$4.9\,pJ/byte), with dynamic power dominating.
The ECC datapath (inner lanes, outer pipes, and differential parity) accounts for about 2.6\,W.
The remaining power comes from channel-facing logic and clocking.
Normalized by throughput, REACH's controller contributes about 4.9\,pJ per byte on the wire, which fits comfortably within a high-end GPU and HBM power budget.

\smallskip
\noindent\textbf{Decoder headroom and latency tails.}
We now quantify how often the outer decoder runs and how it shapes latency.
For BER $=10^{-3}$ and the LLM-style access mix from Section~\ref{sec:q2-service} (95\% reads, 5\% writes, 95\% sequential, 5\% random, differential parity on writes), the measured outer escalation probability is $p_{\text{outer}}\approx 2.4\times 10^{-3}$ per request.
With 26 outer pipes and a fixed 32-cycle repair pipeline at 1.74GHz, the outer cluster runs at about 20\% utilization at 3.35\,TB/s, leaving headroom for higher BER or more random access.

The inner RS path is 12 stages, or about 6.9ns at 1.74GHz.
Requests that trigger an outer repair take 37 cycles, or about 21.3ns.
Table~\ref{tab:latency-tails} reports the ECC service latency distribution when we sample requests according to $p_{\text{outer}}$ and ignore queuing.
Almost all requests complete on the inner path. The outer repair appears only in the far tail, which matches the quantitative study in Section~\ref{sec:verification}.

\vspace{-10px}
\begin{table}[htbp]
    \centering
    \caption{ECC service latency distribution (no queuing) at BER=$10^{-3}$. Inner RS repairs dominate; the outer RS path appears only in the extreme tail.}
    \label{tab:latency-tails}
    \begin{tabular}{lcc}
        \hline
        Percentile & Latency [ns] & Dominant path \\
        \hline
        p50   &  6.90 & Inner only \\
        p90   &  7.03 & Inner only \\
        p99   &  7.21 & Inner only \\
        p99.9 & 21.27 & Inner + outer \\
        \hline
    \end{tabular}
\end{table}

\smallskip
\noindent\textbf{Comparison to a naive long-RS baseline.}
We compare two designs synthesized with the same flow and PDK: (1) a naive long-RS baseline that applies only a wide outer RS decoder to every 2\,KB codeword, and (2) REACH (the two-level design with inner RS and erasure-only outer RS). Table~\ref{tab:ppa-baseline} reports PPA at BER=$10^{-3}$ and a 3.35\,TB/s HBM link.

\begin{table*}[htbp]
    \centering
    \caption{PPA comparison of a naive long-RS baseline and REACH at BER=$10^{-3}$. Both target a 3.35\,TB/s HBM interface using the same ASAP7-based flow.}
    \label{tab:ppa-baseline}
    \begin{tabular}{lccccc}
        \hline
        Design                    & Freq [GHz] & Thru [GB/s] & ECC area [mm$^2$] & ECC power [W] & Outer pipes \\
        \hline
        Naive outer-only RS       & 1.69       & 3458        & 176.7             & 44.5          & 20744    \\
        REACH & 1.74       & 3562        & 15.2              & 17.5          & 26          \\
        \hline
    \end{tabular}
\end{table*}

The naive design reaches the throughput target only by instantiating about $2.1\times 10^4$ full locator pipes, which consumes 176.7mm$^2$ and 44.5\,W for ECC alone. This scale is not realistic for a GPU-class chip. REACH meets the same throughput target with 15.2\,mm$^2$ and 17.5\,W, reducing ECC area by 11.6$\times$ and ECC power by about 60\%, and shrinking the long-code cluster from 20744 full pipes to 26 erasure-only pipes.
This turns the long RS decoder from a monolithic throughput bottleneck into a small, low-duty-cycle repair engine.

\subsection{Importance-Adaptive ECC}
\begin{figure}[htbp]
    \centering
    \includegraphics[width=\linewidth]{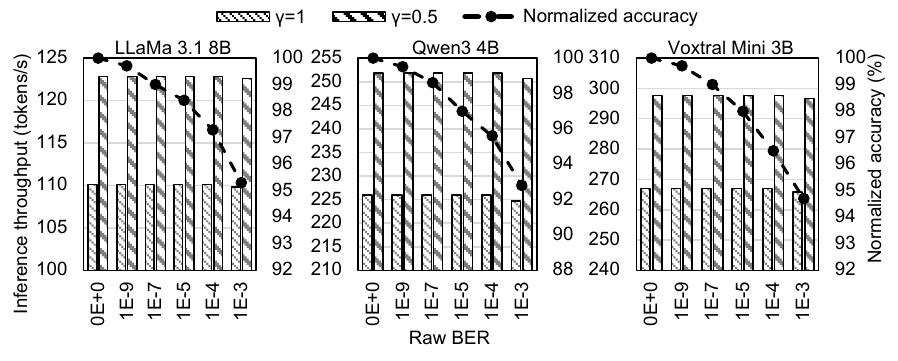}
    \caption{Inference throughput and normalized accuracy on PIQA of importance-adaptive ECC with bit-plane protection.}
    \label{fig:importance_adaptive}
\end{figure}

As shown in Fig.~\ref{fig:importance_adaptive}, we treat reliability as a budgeted resource and protect only the bit-planes that matter as discussed previously. Unprotected planes carry no inner short-RS and no outer long-RS, they move as plain 32~B chunks. Let $\gamma$ be the protected-plane ratio. A setting of $\gamma{=}1.0$ protects all planes, while $\gamma{=}0.5$ protects only the critical planes (e.g., exponents in BF16/FP8). This knob reduces both wire traffic and parity capacity in direct proportion to the share of protected planes. 

We observe a consistent throughput gain when we reduce $\gamma$ from 1.0 to 0.5. At BER$=0$, tokens/s rise from 110.1 to 122.8 for LLaMA-3.1-8B (+11.6\%), from 226.0 to 251.8 for Qwen3-4B (+11.4\%), and from 267.0 to 297.7 for Voxtral-Mini-3B (+11.5\%). The gain persists at high BER because fewer protected planes means fewer escalations and fewer parity bytes in flight. These improvements directly reflect the reduced wire amplification and the lower load on the outer decoder. We also quantify normalized accuracy for PIQA shown in Fig.~\ref{fig:importance_adaptive}. Accuracy remains near baseline at low BER and degrades as BER increases. For example, for LLaMA-3.1-8B, the normalized accuracies of applying $\gamma$=0.5 are 99.7\%, 99.0\%, 98.4\%, 97.3\%, and 95.3\% for BER values $\{10^{-9},10^{-7},10^{-5},10^{-4},10^{-3}\}$.

\vspace{-15px}
\section{Related Works and Discussion}
\label{sec:related}

\noindent\textbf{DRAM scaling and system-level protection.}
As DRAM process nodes scale, reliability barriers such as Variable Retention Time (VRT), RowHammer, and single-event upsets have intensified~\cite{liu2013experimental, kim2014flipping}. To mitigate these, the industry has adopted a layered approach. At the device level, DDR5 and HBM integrated On-die ECC and link protections (e.g., DBI, CRC) to mask single-bit errors locally~\cite{jedec-jesd79-5d, JEDEC_JESD238B01_2025, gurumurthi2021hbm3, synopsys_hbm3,nair2016xed}. However, these device-fixed features lock the code length and redundancy policy into the stack, forcing tight binning that inflates \$/GB compared to commodity DRAM~\cite{Koch2024TheMW, 11185353}. 
At the system level, server platforms employ rank-level protections like Chipkill, SDDC, or virtualization-based retirement~\cite{dell1997white, nvidia_row_remapping2023,udipi2012lot,yoon2010virtualized}. Recent CXL-attached memory work revisits controller-side cross-channel ECC, with costly parity updates on writes~\cite{liu2025cxl}. While effective against catastrophic chip failures, these schemes rely on wide, synchronized rank accesses that are incompatible with HBM’s independent channel architecture and unique error patterns (e.g., TSV faults)~\cite{wu2024removing}. REACH complements these studies by targeting the gap between high raw defectivity and system integrity, moving beyond simple masking to active yield expansion.

\smallskip
\noindent\textbf{Controller-centric ECC evolution.}
Storage systems offer a precedent for handling high error rates via the controller. As NAND flash scaled, controllers evolved from simple BCH codes to sophisticated LDPC schemes with soft-decision decoding and DSP techniques~\cite{sun2006use, zhao2013ldpc, dong2013enabling, li2016elastic, zhang2016real}. Recent work has also explored integrating stronger ECC into DRAM controllers for PCM~\cite{qureshi2009enhancing}. However, blindly porting storage-class codes to HBM fails due to two constraints: (1) \textit{Latency:} Iterative LDPC decoding is too slow for the nanosecond-scale access times of main memory; (2) \textit{Granularity:} Storage operates on large 4\,KB+ pages, whereas HBM enforces 32\,B transactions. Prior logic-layer integration approaches (e.g., HMC or 3D-stacked logic) allowed custom protocols but failed to gain traction against JEDEC standards~\cite{pawlowski2011hybrid,jeddeloh2012hybrid,ahn2015scalable}. REACH adapts the controller-centric philosophy to the rigid constraints of the HBM interface by introducing the erasure-only decoding path, effectively reconciling strong protection with low latency.

\smallskip
\noindent\textbf{Generalizability and alternative architectures.}
A natural alternative design space includes stronger controller-side codes (e.g., BCH, LDPC, or Chipkill-like symbols). We do not claim that RS code is the only viable option. However, long-span codes operating over fixed-granularity interfaces, whether HBM, DDR, LPDDR, or CXL-attached memory, must contend with two structural challenges: (i) read--modify--write amplification on small random accesses, and (ii) the area and power scaling of error-locator or iterative-decoding logic at multi-TB/s bandwidths. REACH addresses these issues by introducing a burst-aligned erasure boundary (e.g., 32\,B for HBM). While our implementation targets HBM for AI inference, this hierarchical erasure-decoding framework is media-agnostic in principle. Commodity server (DDR5) and mobile (LPDDR5) memories face similar scaling pressures, such as variable retention time and RowHammer. REACH adapts to these standards by aligning the inner code with their native burst granularity (e.g., 64B for DDR), offering a generalizable framework for decoupling raw device reliability from system integrity.

\section{Conclusion}
\label{sec:conclusion}

HBM cost per GB is now a first-order limiter for LLM inference, and fixed on-die ECC forces short codes and tight binning, raising price and blocking workload-aware tuning. This paper demonstrates that raw HBM reliability can be relaxed at the device level without compromising system integrity. We presented REACH, a controller-centric architecture that preserves the standard HBM PHY. It employs an inner per-32\,B code to verify and locally correct most faults and flag only hard chunks as erasures, and a long outer RS that runs in erasure-only mode. Differential parity bounds small-access amplification, and an importance-adaptive bit-plane layout protects only critical fields. Evaluated on LLaMA-3.1-8B, Voxtral-Mini-3B, and Qwen3-4B at 8K context, the design sustains about 79\% of on-die ECC throughput at BER=0 (error free), remains qualified up to raw BER $\sim 10^{-3}$, and in ASAP7 occupies 15.2\,mm$^2$ and 17.5\,W at 3.56\,TB/s while reducing ECC area and power by 11.6$\times$ and $\sim$60\% versus a naive long-RS baseline. These results demonstrate that long-code ECC can be treated as a tunable \emph{system} resource. By co-designing inner/outer codes, parity math, and data layout in the controller, REACH preserves HBM-class throughput, substantially relaxes device-level raw reliability targets, and opens a practical path to lower HBM cost per bit for AI inference.


\bibliographystyle{ACM-Reference-Format}
\bibliography{ISCA2026_IEEE_template/refs}

\end{document}